\documentclass[12pt]{article}
\usepackage{epsfig,amssymb,amsmath,psfrag}

\usepackage{pstricks}
\usepackage{color}
\usepackage{accents}

\def \be  {\begin{equation}}
\def \ee  {\end{equation}}
\def \ba  {\begin{eqnarray}}
\def \ea  {\end{eqnarray}}

\def \sha{{\,\amalg\hskip -3.6pt\amalg\,}}

\begin{document}

\thispagestyle{empty}

\null\vskip-12pt \hfill  CERN-PH-TH/2012-363\\
\null\vskip-12pt \hfill  LAPTH-062/12 \\

\vskip2.2truecm
\begin{center}
\vskip 0.2truecm {\Large\bf
{\Large Superstring amplitudes and the associator}
}\\
\vskip 1truecm
{\bf J.~M. Drummond${}^{1,2}$ and E. Ragoucy${}^{2}$ \\
}

\vskip 0.4truecm
{\it
${}^{1}$ CERN, Geneva 23, Switzerland\\
\vskip .2truecm                        }
\vskip .2truecm
{\it
${}^{2}$ LAPTH, Universit\'{e} de Savoie, CNRS\\
B.P. 110,  F-74941 Annecy-le-Vieux Cedex, France\\
\vskip .2truecm                        }
\end{center}

\vskip 1truecm 
\centerline{\bf Abstract} 

We investigate a pattern in the $\alpha'$ expansion of tree-level open superstring amplitudes which correlates the appearance of higher depth multiple zeta values with that of simple zeta values in a particular way. We rephrase this relationship in terms of the coaction on motivic multiple zeta values and show that the pattern takes a very simple form, which can be simply explained by relating the amplitudes to the Drinfel'd associator derived from the Knizhnik-Zamolodchikov equation. Given this correspondence we show that, at least in the simplest case of the four-point amplitude, the associator can be used to extract the form of the amplitude.

\medskip

 \noindent

\newpage
\setcounter{page}{1}\setcounter{footnote}{0}


\section{Introduction and summary}

Much recent progress in understanding the structure of scattering amplitudes in field theory has been based on understanding the analytic structure of loop corrections in terms of the Hopf structure underlying the iterated integrals defining multiple polylogarithms. Results along these lines have been obtained in \cite{Goncharov:2010jf,Gaiotto:2011dt,Dixon:2011pw,Heslop:2011hv,Dixon:2011nj,Duhr:2012fh}, primarily based on using the symbol \cite{Chen,FBthesis,Gonch} or motivic coaction \cite{Gonch2,Brown:2011ik,fb2} as a tool to perform analysis of the analytic structure in an algebraic manner. While loop corrections in field theory quickly become rather complicated, there is a similar problem in which a transcendental structure appears in a much milder form, namely string theory scattering amplitudes at tree level. If one expands tree-level string amplitudes in $\alpha'$, the inverse string tension, one finds coefficients which contain multiple zeta values, rather than multiple polylogarithms.

In \cite{Schlotterer:2012ny} an intriguing pattern of multi-zeta values was found in the $\alpha'$ expansion of the open superstring amplitudes. The coefficients of all multiple zeta values of depth greater than one are fixed in terms of those of depth one in a specific way, indicating that in the case of superstring amplitudes, the Hopf structure is not just a tool that can be used to analyse the results, but rather it is intrinsic to their structure.

The pattern among the multiple zeta values was expressed in \cite{Schlotterer:2012ny} in terms of an auxiliary map $\phi$ which relates the (motivic) multi-zetas to a Hopf algebra of words endowed with the shuffle product. The map was introduced in \cite{Brown:2011ik,fb2} to help to explain the structure of the space of motivic multi-zetas and to produce an efficient algorithm for fixing a basis and projecting a given multi-zeta value into that basis.
As noted in \cite{Schlotterer:2012ny} this pattern is closely related to a representation of the identity operator on the Hopf algebra of words, so that the pattern is something canonical. However the objects introduced to encode the structure do depend on the choice of basis of multiple zeta values. In particular, the isomorphism $\phi$ between the multi-zetas and the Hopf algebra of words is non-canonical in that it depends on the basis of multi-zetas chosen. One of the aims of the present work is to express this pattern purely in terms of the Hopf structure associated to multi-zeta values, without introducing a basis.

Indeed we will see that we can rephrase the pattern found in  \cite{Schlotterer:2012ny} purely in terms of the coaction on motivic multiple zetas, making the independence on the choice of basis manifest. Following \cite{Mafra:2011nv,Mafra:2011nw} we express the colour-ordered open superstring amplitudes in terms of a matrix $R$ acting on a vector of independent colour-ordered super Yang-Mills amplitudes,
\be
A^{\rm open} = R A^{\rm YM}\,.
\ee
The matrix $R = 1\!\!1 + O (\alpha ')$ encodes all the $\alpha'$ corrections to the super Yang-Mills amplitudes. In terms of the motivic coaction, the pattern found in \cite{Schlotterer:2012ny} can be rephrased as follows,
\be
\Delta R^{\rm m} = R^{\rm m} \dot{\otimes} R^{\rm a}\,.
\label{DeltaR}
\ee
The superscripts on $R$ refer to the fact that all multi-zeta values should be replaced by their motivic versions (so that the coaction is defined) and moreover that in the right-hand factor we should work modulo $\zeta_2^{\rm m}$. The equation (\ref{DeltaR}) implies that all coefficients of multiple zetas of depth greater than one are fixed in terms of those of depth one, in complete agreement with the structure presented in \cite{Schlotterer:2012ny}.

Moreover the equation (\ref{DeltaR}) is similar to a property obeyed by another object, the Drinfel'd associator $\Phi$ \cite{Drinfeld:1989st,Drinfeld2}, defined in terms of the Knizhnik-Zamolodchikov equation \cite{Knizhnik:1984nr}. It is a form of universal monodromy for solutions of the KZ equation. The associator can be written as a generating series for all (shuffle regularised) multiple zeta values,
\be
\Phi = \sum_w w \zeta_{\sha} (w)\,,
\ee
where the $w$ are words in two letters. $\Phi$ obeys the relation
\be
\Delta \Phi^{\rm m} = \Phi^{\rm m} \accentset{\triangleleft}{\otimes} \Phi^{\rm a}\,,
\ee
in complete analogy with (\ref{DeltaR}), where $\triangleleft$ is the Ihara action \cite{Ihara,DG,fbdg} on group-like series of words.

Indeed we will see in the simplest case (the four-point amplitude) that we can actually identify the associator $\Phi$ with the matrix $R$ appearing in the open superstring amplitude, which actually allows one to fix also the coefficients of the zeta values of depth one.

The paper is organised as follows. We begin with a very brief review of superstring amplitudes in section \ref{sect-ssamps}. We introduce multiple zeta values and the motivic coaction in section \ref{sect-MZVs}. Then, in section \ref{sect-patterns} we describe the patterns in the $\alpha'$ expansion of the amplitudes found in \cite{Schlotterer:2012ny} and describe how they can be rephrased as in eq (\ref{DeltaR}). In section \ref{sect-associator} we introduce the associator and describe its properties under the motivic coaction. In section \ref{sect-KZ} we use the Knizhnik-Zamolodchikov equation to derive the form of the open superstring four-point amplitude. In section \ref{sect-closed} we make some remarks on the structure of closed superstring amplitudes and the constraints on the multiple zeta values appearing in both closed and open superstring amplitudes.

\section{Tree-level open superstring scattering amplitudes}
\label{sect-ssamps}

Let us begin with tree-level scattering amplitudes in gauge theory. The $n$-gluon amplitude of a gauge theory can be decomposed as a sum over cyclic colour-ordered partial amplitudes,
\be
\mathbb{A}^{\rm gauge} = \sum_{\sigma \in S_n / C_n} A^{\rm gauge}(\sigma(1),\ldots,\sigma(n)) {\rm Tr}(T^{a_{\sigma(1)}} \ldots T^{a_{\sigma(n)}})\,.
\ee
The tree-level open superstring amplitude $\mathbb{A}^{\rm open}$ of massless gauge bosons can similarly be decomposed in terms of colour-ordered partial amplitudes $A^{\rm open}$. The superstring amplitudes depend on $\alpha'$, the inverse string tension and in the limit $\alpha'\rightarrow 0$ one recovers the gauge theory amplitudes. All of these statements are independent of the number of uncompactified dimensions in which one studies the scattering process.

The colour-ordered partial amplitudes in gauge theory obey certain relations. The simplest among these are the cyclic and reflection identities. Then one has the photon decoupling identity and Kleiss-Kuijf relations \cite{Kleiss:1988ne}. Finally there are the BCJ relations \cite{Bern:2008qj}. These relations among gauge theory partial amplitudes can be derived from the monodromy properties of the open string theory partial amplitudes \cite{BjerrumBohr:2009rd,Stieberger:2009hq,BjerrumBohr:2010zs}. If one uses all these relations to reduce the set of colour-ordered partial amplitudes to a minimal set, then $(n-3)!$ partial amplitudes remain. One may choose these partial amplitudes to be those obtained from all permutations of the labels $2,\ldots,n-2$, keeping $1, (n-1)$ and $n$ fixed. We may arrange these remaining amplitudes into a vector which we denote by $A^{\rm gauge}$, or correspondingly, $A^{\rm open}$.

Thus we have a vector of independent colour-ordered open superstring amplitudes related to the corresponding field theory amplitudes via a matrix $R$,
\be
A^{\rm open} = R A^{\rm gauge}\,.
\ee
The matrix $R$ has an expansion $R = 1\!\!1 + O(\alpha')$. In general, terms of order $(\alpha')^m$ contain multi-zeta values of weight $m$. The detailed structure of the $\alpha'$ corrections has been studied in \cite{Mafra:2011nv,Mafra:2011nw}. As one might expect it is given in terms of $(n-3)$-fold integrals over the positions of the vertex operators not fixed using conformal symmetry. Here we will outline the simplest cases $n=4,5$.

The simplest case is $n=4$ where the matrix $R$ is therefore a $(1 \times 1)$ matrix.
In this case one has \cite{Green:1981xx} 
\be
R = \frac{\Gamma(1-s)\Gamma(1-t)}{\Gamma(1-s-t)} \,.
\ee
Here $s = \alpha' (p_1 + p_2)^2$ and $t = \alpha' (p_2+p_3)^2$ are the two Mandelstam variables and we recall that the momenta satisfy $p_i^2=0$ since we are considering the on-shell scattering of massless gauge bosons.
If we expand the gamma functions as a series in $s$ and $t$ we find zeta values. Specifically, we can write $R$ as the exponential of an infinite series of contributions proportional to simple zeta values $\zeta_n=\zeta(n)$, 
\be
R = {\rm exp} \sum_{n \geq2}  \frac{\zeta_n}{n} [s^n +t^n - (s+t)^n]\,.
\ee
If we wish we may decompose the exponential into the contributions of even and odd zeta values,
\be
R = P E\,
\ee
where
\be
P ={\rm exp} \sum_{n \geq 1} \frac{\zeta_{2n}}{2n} [s^{2n} + t^{2n} - (s+t)^{2n}]\,
\ee
and
\be
E = \exp \sum_{n \geq1} \frac{\zeta_{2n+1}}{2n+1} [s^{2n+1} + t^{2n+1} - (s+t)^{2n+1}]\,.
\ee

For five-point amplitudes, $R$ is a $(2 \times 2)$ matrix,
\be
R = 
\left(
\begin{matrix}
F_1 & F_2 \\
\tilde{F}_2 & \tilde{F}_1
\end{matrix}
\right)
\ee
where we have 
\begin{align}
F_1 &= s_{12} s_{34} \int_0^1 dx \int_0^1 dy x^{s_{45}} y^{s_{12} -1} (1-x)^{s_{34}-1} (1-y)^{s_{23}} (1-xy)^{s_{24}}\,, \notag \\
F_2 &= s_{13} s_{24} \int_0^1 dx \int_0^1 dy x^{s_{45}} y^{s_{12} } (1-x)^{s_{34}} (1-y)^{s_{23}} (1-xy)^{s_{24}-1}\,.
\end{align}
The five independent Mandelstam variables are chosen to be $s_{12}, s_{23}, s_{34}, s_{45}, s_{24}$ with $s_{ij} = (p_i + p_j)^2$. The functions $\tilde{F}_1$ and $\tilde{F}_2$ are given by $\tilde{F}_i = F_i|_{p_2 \leftrightarrow p_3}$. The integrals above can be expressed in terms of hypergeometric functions and gamma functions \cite{Kitazawa:1987xj,Mafra:2011nw}. 
\begin{align}
\frac{F_1}{ s_{12} s_{34}} = &\frac{\Gamma(1 + s_{45}) \Gamma(s_{12}) \Gamma(s_{34}) \Gamma(1 + s_{23})}{\Gamma(1+s_{45}+s_{34})\Gamma(1+s_{12}+s_{23})} \notag \\
&\times {}_3F_2(1 + s_{45}, s_{12}, -s_{24}; 1 + s_{45} + s_{34}, 1 + s_{12} + s_{23}; 1) \,,\\
\frac{F_2}{s_{13}s_{24}}= &\frac{\Gamma(1 + s_{12}) \Gamma(1+s_{23}) \Gamma(1+s_{34}) \Gamma(1 + s_{45})}{\Gamma(2+s_{12}+s_{23})\Gamma(2+s_{34}+s_{45})} \,,\notag \\
&\times {}_3F_2(1 + s_{12},1+ s_{45},1 -s_{24}; 2 + s_{12} + s_{23}, 2 + s_{34} + s_{45}; 1)\,.
\end{align}

If we expand the above hypergeometric functions as a series in the Mandelstam variables we find multiple zeta values. Before discussing the patterns in the expansion found in \cite{Schlotterer:2012ny} we will give a short introduction to multiple zeta values and the Hopf structure of their motivic counterparts.

\section{Multi-zeta values and their motivic counterparts}
\label{sect-MZVs}

Multi-zeta values can be defined in terms of nested sums. Here we will use an iterated integral representation. 
We consider iterated integrals of the following form,
\be
I(a_0;a_1,\ldots ,a_n;a_{n+1}) = \int_\gamma \frac{dz_1}{z_1-a_1} \ldots \frac{dz_n}{z_n-a_n}
\ee
where $\gamma$ is a path from $a_0$ to $a_{n+1}$ avoiding the poles at $a_1,\ldots,a_n$, and the integral is performed so that $a_0,z_1,\ldots,z_n,a_{n+1}$ gives an ordering along the curve. If the poles coincide with the endpoints the integral may require regularisation.

Sometimes we will use the notation $I(a_0;w;a_1)$ where $w$ is the word $a_1 \ldots a_n$.
The multi-zeta values are special cases where $a_0=0$ and $a_{n+1}=1$, with $\gamma$ being the path along the real axis, and $a_i \in \{ 0,1 \}$ for $1\leq i \leq n$. There is a also a sign depending on the number $r$ of the $a_i$ equal to 1,
\be
\zeta(p_1,\ldots,p_r) = (-1)^r I(0;10^{p_1-1} \ldots 1 0^{p_r-1};1)\,. 
\ee
We take $p_r \geq 2$ to ensure convergence of the integral. Very often we write $\zeta_{p_1,\ldots,p_r}$ to save a little space.
Even more compactly we can write
\be
\zeta(w) = (-1)^r I(0;w;1)
\ee
where $w$ is a word built from the letters $\{ 0,1 \}$, beginning with a 1 and ending with a 0, and $r$ is the number of ones in $w$ as before.
The simplest example of a zeta value is given by
\be
\zeta_2= - I(0;10;1)  = - \int_0^1 \frac{dt_1}{t_1-1} \int_{t_1}^{1} \frac{dt_2}{t_2}\,.
\ee
All even simple zeta values are rational multiples of powers of $\zeta_2$,
\be
\zeta_{2n} = b_n \zeta_2^n\,,\qquad b_n = (-1)^{n+1}\frac{1}{2} B_{2n} \frac{(24)^n}{(2n)!}\,,
\label{evenzetas}
\ee
where $B_{2n}$ denotes the Bernoulli numbers.

The iterated integrals obey a shuffle product relation,
\be
I(a_0;w_1;a_{n+1}) I(a_0;w_2;a_{n+1}) = I(a_0; w_1 \sha w_2 ; a_{n+1})\,,
\label{Ishuff}
\ee
where we recall that the shuffle product of two words $w_1 \sha w_2$ is a sum over all words given by permutations of the elements of $w_1$ and $w_2$ which preserve the orderings within $w_1$ and $w_2$.
For $a_0=0$ and $a_{n+1}=1$ this implies the shuffle relation for the multiple zeta values,
\be
\zeta(w_1) \zeta(w_2) = \zeta(w_1 \sha w_2)\,.
\ee
We can use the above relation to define a regularised version of the multi-zeta values whose words begin with 0 or end in a 1; we extract any leading zeros and trailing ones from a given zeta using the shuffle relation and define $\zeta_\sha(0)=\zeta_\sha(1) = 0$.

One may lift iterated integrals $I$ to their `motivic' versions $I^{\rm m}$. These are defined as abstract elements of an algebra $\mathcal{H}^{\mathcal{MT}}$, graded by weight and obeying the shuffle product relation (\ref{Ishuff}). The motivic iterated integrals satisfy all known algebraic relations satisfied by real iterated integrals and conjecturally capture all such relations. As special cases of the motivic iterated integrals one has the motivic multiple zeta values $\zeta^{\rm m}(w) = I^{\rm m}(0;w;1)$.

Since all even simple zeta values are known to be rational multiples of powers of $\zeta_2$, the element $\zeta^{\rm m}_2$ plays a special role. It is convenient to introduce the quotient $\mathcal{A}^{\mathcal{MT}} = \mathcal{H}^{\mathcal{MT}} / \zeta^{\rm m}_2$, whose elements are denoted $I^{\rm a}$ (or $\zeta^{\rm a}$ when specialising to motivic MZVs). Then $\mathcal{A}^{\mathcal{MT}}$ is a Hopf algebra and $\mathcal{H}^{\mathcal{MT}}$ is a trivial comodule over $\mathcal{A}^{\mathcal{MT}}$ which can be non-canonically identified with $\mathbb{Q}[\zeta_2^{\rm m}]\otimes \mathcal{A}^{\mathcal{MT}}$. As an algebra it is commutative, with the product being the shuffle product (\ref{Ishuff}). The coaction $\Delta : \mathcal{H}^{\mathcal{MT}} \longrightarrow \mathcal{H}^{\mathcal{MT}} \otimes \mathcal{A}^{\mathcal{MT}}$ is defined on the motivic iterated integrals as follows \cite{Gonch2,Brown:2011ik,fb2}. 
\begin{align}
&\Delta I^{\rm m}(a_0;a_1,\ldots,a_n;a_{n+1}) = \notag \\
&\sum_k \sum_{i_0 < i_1  \ldots i_k <i_{k+1}} I^{\rm m}(a_0 ; a_{i_1} , \ldots , a_{i_k} ; a_{n+1}) \otimes \Bigl( \prod_{p=0}^k I^{\rm a}(a_{i_p};a_{i_p +1}, \ldots , a_{i_{p+1}-1} ; a_{i_{p+1}}) \Bigr) \,.
\label{coproduct}
\end{align}
The terms in the above formula can be associated with polygons inscribed into the semi-circle with $(n+1)$ marked points $a_j$. Note we have reversed the order of the factors on the RHS with respect to \cite{Brown:2011ik,fb2}. The above coaction specialises in the obvious way to the algebra of motivic multiple zeta values, denoted by $\mathcal{H}$.
\begin{figure}
\psfrag{a0}[cc][cc]{$a_0$}
\psfrag{ai1}[cc][cc]{$a_{i_1}$}
\psfrag{ai2}[cc][cc]{$a_{i_2}$}
\psfrag{ai3}[cc][cc]{$a_{i_3}$}
\psfrag{ai4}[cc][cc]{$a_{i_4}$}
\psfrag{anp1}[cc][cc]{$a_{n+1}$}
\centerline{{\epsfysize6cm \epsfbox{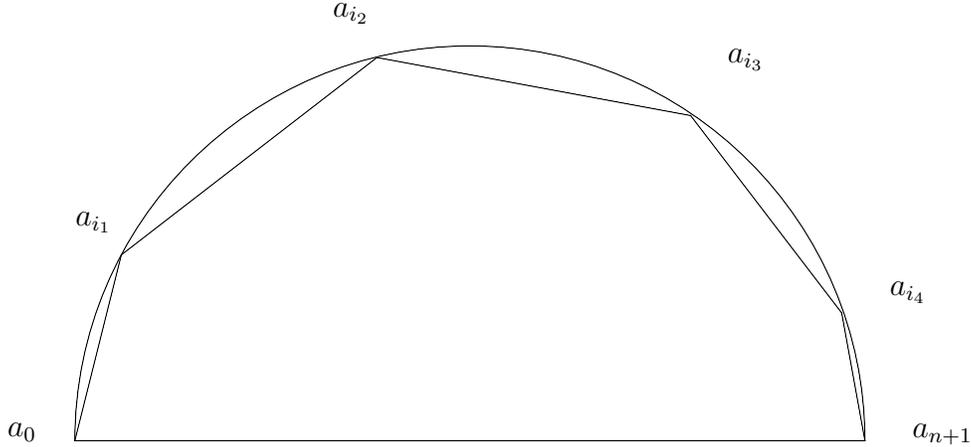}}} \caption[]{\small A contribution to the motivic coaction for $k=4$}
\label{Fig:aux}
\end{figure}

The only primitive elements of $\mathcal{H}$ are the simple zeta values $\zeta^{\rm m}_{n}$. Because of the fact that one should mod out by $\zeta^{\rm m}_2$ on the right-hand factor, the coaction looks different for even and odd simple zetas,
\be
\Delta \zeta^{\rm m}_{2n} = \zeta^{\rm m}_{2n} \otimes 1, \qquad \Delta \zeta^{\rm m}_{2n+1} = \zeta^{\rm m}_{2n+1} \otimes 1 + 1 \otimes \zeta^{\rm a}_{2n+1}\,.
\ee

\section{Patterns in the $\alpha'$ expansion}
\label{sect-patterns}

In \cite{Schlotterer:2012ny} the following pattern for $R$ in the five-point case was found up to weight 16,
\be
R = P Q E\,
\ee
where 
\be
P = 1+ \sum \zeta_{2n} P_{2n}\,,
\ee
\be
E = : \exp \sum \zeta_{2n+1} M_{2n+1} :
\ee
and
\be
Q = 1+ \sum_{n\geq 8} Q_n\,.
\ee
The normal ordering symbol in $E$ means that $:M_i M_j: = M_j M_i$ if $i<j$ and $:M_i M_j: = M_i M_j$ otherwise. Alternatively one may write $E$ as an ordered product of exponentials $E= (\ldots e^{\zeta_5 M_5} e^{\zeta_3 M_3})$. The matrices $Q_n$ contain multiple zeta values of higher depth and are determined in terms of commutators of the $M_{2n+1}$. Thus if all the $M_n$ commute with each other, as is the case for the four-point amplitude, the matrix $Q$ reduces to the identity. The structure is therefore a generalisation of the one for the four-point amplitude to non-commuting $M_i$. 

Before we discuss in detail the structure from a Hopf algebraic point of view, let us exhibit the first few terms in $Q$ found in \cite{Schlotterer:2012ny} in order to illustrate how the coefficients of multiple zetas of higher depth are fixed in terms of those of depth 1. For this we need to make our basis of multi zeta values explicit. Up to weight 12 we take $\{\zeta_2,\zeta_3,\zeta_5,\zeta_7,\zeta_{3,5},\zeta_9,\zeta_{3,7},\zeta_{11},\zeta_{3,3,5},\zeta_{3,9},\zeta_{1,1,4,6}\}$ and all possible products of these elements. In this basis the $Q_n$ found in \cite{Schlotterer:2012ny} take the form,
\begin{align}
Q_8 &=  \frac{1}{5} \zeta_{3,5} [M_5,M_3]\,,\\
Q_9 &= 0\,, \\
Q_{10} &= \biggl(\frac{3}{14} \zeta_5^2 + \frac{1}{14} \zeta_{3,7}\biggr)[M_7,M_3]\,,\\
Q_{11} &= \biggl(9 \zeta_2 \zeta_9 + \frac{6}{25} \zeta_2^2 \zeta_7 - \frac{4}{35} \zeta_2^3 \zeta_5 + \frac{1}{5} \zeta_{3,3,5}\biggr)[M_3,[M_5,M_3]]\,, \\
Q_{12} &= \bigl( \frac{2}{9} \zeta_5 \zeta_7 + \frac{1}{27} \zeta_{3,9}\biggr)[M_9,M_3] \notag \\
&\quad + \frac{48}{691}\biggl(\frac{18}{35} \zeta_2^3 \zeta_3^2 + \frac{1}{5} \zeta_2^2 \zeta_3 \zeta_5 - 10 \zeta_2 \zeta_3 \zeta_7 -\frac{7}{2} \zeta_2 \zeta_5^2 - \frac{3}{5} \zeta_2^2 \zeta_{3,5}  - 3 \zeta_2 \zeta_{3,7} \notag\\ 
 \quad& - \frac{1}{12} \zeta_3^4 - \frac{467}{108} \zeta_5 \zeta_7 + \frac{799}{72} \zeta_3 \zeta_9 + \frac{2665}{648} \zeta_{3,9} + \zeta_{1,1,4,6}\biggr) \bigl([M_9,M_3] - 3 [M_7,M_5]\bigr)
\end{align}

To be completely explicit, let us expand the matrix $R$ up to order $(\alpha')^{10}$,
\begin{align}
R &= 1 + \zeta_2 P_2 + \zeta_3 M_3 + \zeta_4 P_4 + \zeta_5 M_5 + \zeta_2 \zeta_3 P_2 M_3 + \frac{1}{2} \zeta_3^2 M_3 M_3 + \zeta_6 P_6 \notag \\
&+ \zeta_7 M_7 + \zeta_2 \zeta_5 P_2  M_5 + \zeta_4 \zeta_3 P_4 M_3  + \frac{1}{5} \zeta_{3,5} [M_5,M_3] +\zeta_5 \zeta_3 M_5  M_3\notag \\
&+ \frac{1}{2} \zeta_2 \zeta_3^2 P_2 M_3 M_3 + \zeta_8 P_8 + \zeta_9 M_9 + \frac{1}{6} \zeta_3^3 M_3 M_3 M_3 +\zeta_2 \zeta_7 P_2 M_7 \notag\\
&+ \zeta_4 \zeta_5 P_4 M_5 + \zeta_6 \zeta_3 P_6  M_3 + \biggl(\frac{3}{14} \zeta_5^2 + \frac{1}{14}\zeta_{3,7}\biggr) [M_7,M_3] + \zeta_7 \zeta_3 M_7 M_3 \notag\\
&+\frac{1}{2} \zeta_5^2 M_5 M_5 +\frac{1}{5} \zeta_2 \zeta_{3,5} P_2 [M_5,M_3] + \zeta_2 \zeta_5 \zeta_3 P_2 M_5 M_3 \notag \\
&+ \frac{1}{2} \zeta_4 \zeta_3^2 P_4 M_3 M_3 + \zeta_{10} P_{10} + \ldots
\label{Rtowt10}
\end{align}
The fact that the coefficients of the higher depth multi zeta values are given in terms of those of the simple zeta values can be explained in terms of the Hopf algebra structure obeyed by the motivic multi zeta values. It is therefore useful to introduce the notation $R^{\rm m}$ to denote the matrix $R$ with the zeta values replaced by their motivic versions and $R^{\rm a}$ to denote the same matrix modulo $\zeta_2^{\rm m}$.

In \cite{Schlotterer:2012ny}, the structure underlying the matrix $R$ was described by introducing an auxiliary Hopf algebra $\mathcal{U}'$, following \cite{Brown:2011ik,fb2}. The Hopf algebra $\mathcal{U'}$ is the commutative, graded Hopf algebra of all non-commutative words in generators of each odd degree $d\geq3$, denoted by $\{f_3,f_5,f_7,f_9,\ldots\}$. These generators play the role of the simple zeta values. The product on $\mathcal{U}'$ is the shuffle product; the coproduct is given by deconcatenation. The even zeta values are taken into account by considering the trivial comodule
\be
\mathcal{U} = \mathbb{Q}[f_2] \otimes \mathcal{U}'\,,
\ee
where $f_2$ is a generator of degree 2. The references \cite{Brown:2011ik,fb2} show how to construct isomorphisms $\phi_B : \mathcal{H} \rightarrow \mathcal{U}$, depending on the choice of basis $B$ of the motivic multi-zeta values $\mathcal{H}$. We describe the isomorphisms $\phi_B$ in more detail in the appendix, describing their basis dependence.

Having introduced the isomorphisms $\phi_B$ associated to a given basis of $\mathcal{H}$, we can now state explicitly the structure found in \cite{Schlotterer:2012ny} for the matrix $R$ describing the open superstring scattering amplitudes. After replacing all zeta values with their motivic versions in the expansion of $R$, one finds the following simple pattern after applying the isomorphism $\phi_B$,
\be
\phi_B (R^{\rm m}) = \biggl(\sum f_2^k P_{2k} \biggr) \sum_p \sum_{i_1,\ldots,i_p}  M_{i_1} \ldots M_{i_p} f_{i_1} \ldots f_{i_p} \,.
\label{phiR}
\ee
We remind the reader that both the matrices $P$ and $M$ and the map $\phi_B$ depend on the choice of basis $B$ for $\mathcal{H}$, although we have not made explicit the dependence of $P_r$ and $M_r$ on $B$. The above structure (\ref{phiR}) for the image of $R$ under $\phi_B$, however, does not depend on $B$. 

Indeed, it was noted in \cite{Schlotterer:2012ny} that (\ref{phiR}) has the same form as the canonical element of $\mathcal{U} \otimes \mathcal{U}^*$. Let us introduce the operators $\partial_{2r+1}$ which act from the right on a word built from the $f_n$ so that $\partial_{2r+1}$ removes the last letter of the word if it is $f_{2r+1}$ or annihilates the word otherwise. Thus $\partial_{r_1} \ldots \partial_{r_n}$ is the dual basis element to $f_{r_1} \ldots f_{r_n}$.
Similarly we can introduce operators $\partial_{2r}$ on the left which pick out the coefficient of $f_{2}^r$. Then replacing $P_{2k}$ by $\partial_{2k}$ and $M_{2r+1}$ by $\partial_{2r+1}$, formula (\ref{phiR}) turns into the canonical element of $\mathcal{U} \otimes \mathcal{U}^*$. Thus the matrices $P_{2k}$ and $M_{2k+1}$ represent the operators $\partial_r$.

In fact one may state the above rather more simply by saying that the matrices $P$ and $M$ represent the duals of the primitive zeta values in a given basis $B$. Thus $R^{\rm m}$ is really representing the canonical element of $\mathcal{H} \otimes \mathcal{H}^*$. At this stage, no information is available on what the representation is, other than by directly expanding the result for the amplitude.

We now observe that the above structure can be restated as the following property of $R$,
\be
\Delta R^{\rm m} = R^{\rm m} \dot{\otimes} R^{\rm a}\,.
\label{coproductrelation}
\ee
Here the symbol $\dot\otimes$ means that the two factors of the tensor product are multiplied as matrices. In addition to its compactness, the above expression makes manifest the fact that the structure found in \cite{Schlotterer:2012ny} is independent of any choice of basis for $\mathcal{H}$. The relation (\ref{coproductrelation}) for the coaction on $R^{\rm m}$ fixes all but the primitive elements, i.e. contributions proportional to $\zeta_n^{\rm m}$, corresponding to the matrices $P_{2n}$ and $M_{2n+1}$ above. Another advantage to (\ref{coproductrelation}) is that, in order to test it, one does not have to refer to the underlying algebra structure of $\mathcal{U}$ or $\mathcal{H}$. One simply computes the motivic coaction on $R^{\rm m}$ and compares with the RHS.

The equation (\ref{coproductrelation}) for $R^{\rm m}$ is equivalent to the fact that $R^{\rm m}$ represents the canonical element $\mathcal{R}$ in $\mathcal{H} \otimes \mathcal{H}^*$.  Indeed we have, in terms of a basis $e_i$ of $\mathcal{H}$ (and $\tilde{e}_k$ of $\mathcal{A}$) and the dual basis $e_i^*$ of $\mathcal{H}^*$ (and $\tilde{e}_k^*$ of $\mathcal{A}^*$),
\begin{align}
\Delta \mathcal{R} &= \Delta \Bigl(\sum_i e_i \otimes e_i^* \Bigr) = \sum_{i,j,k} f_{jki} (e_j \otimes \tilde{e}_k) \otimes e_i^* \notag \\
 &= \mu \sum_j (e_j \otimes e_j^*) \otimes \sum_k (\tilde{e}_k \otimes \tilde{e}_k^*) = \mathcal{R} \dot{\otimes} \tilde{\mathcal{R}}.
\end{align}
Here $\Delta e_i = \sum_{j,k} f_{jki} e_j \otimes \tilde{e}_k$ describes the coaction on $\mathcal{H}$ and $\mu (e_j^* \otimes \tilde{e}_k^*) = \sum_i f_{jki} e_i^*$ describes the action on the duals. We have checked explicitly that the expressions given in \cite{Schlotterer:2012ny} do indeed satisfy the relation (\ref{coproductrelation}) to weight 15.

The consequence of (\ref{coproductrelation}) is that the amplitude is completely determined in terms of the matrices $P_{2n}$ and $M_{2n+1}$. The matrices $P_{2n}$ and $M_{2n+1}$ themselves are not determined. This is already clear at the level of the four-point amplitude since any expression of the form
\be
R^{\rm m} = {\rm exp}\Bigl\{ \sum c_n \zeta_n^{\rm m} \Bigr\}
\ee
will obey $\Delta R^{\rm m} = R^{\rm m} \otimes R^{\rm a}$. One may ask if there is any connection to the Hopf structure of zeta values which fixes the coefficients $c_n$ to be precisely those appearing in the four-point amplitude.

\section{Connection to the Drinfel'd associator}
\label{sect-associator}

By comparing formulas appearing in \cite{Schlotterer:2012ny} and \cite{fbdg} one can see that the structure of the matrix $R$ describing the open superstring amplitude mirrors closely the structure of another object, the Drinfel'd associator \cite{Drinfeld:1989st,Drinfeld2}. This is a form of universal monodromy for the Knizhnik-Zamolodchikov equation.

The associator can be written as a generating function for all multi-zeta values,
\be
\Phi = \sum_w w \zeta_\sha(w)\,,
\label{defassoc}
\ee
where the sum is over all words $w$ in the alphabet $\{e_0 , e_1\}$ and the multi-zeta values are regularised using the shuffle product. The space of words on $\{e_0,e_1\}$ can be identified with the universal enveloping algebra $U(g)$ of the free Lie algebra on two generators $g={\rm Lie}[e_0,e_1]$\,. Thus $\Phi$ is an element of $U(g)$ over the reals.

To express the associator it is useful to introduce derivations acting on $U(g)$ as follows. Given an element $y$ of $g$ we define (following \cite{Ihara1,Ihara})
\be
D_y e_0 = 0, \qquad D_y e_1 = [e_1,y]\,,
\ee
and extend $D_y$ as a derivation to the whole of $U(g)$.
Given the derivations $D$ above we can define a right action of $g$ on $U(g)$ as follows,
\be
x \circ y = x y - D_y x, \qquad y \in g, \quad x \in U(g)\,.
\ee
The Ihara bracket is defined as its antisymmetrisation after restricting $U(g)$ to $g$ in the natural way,
\be
\{ x , y \} = x \circ y - y \circ x = [x,y] +D_x y - D_y x\,, \qquad x,y \in g\,.
\ee
Note that $\{x,y\}$ is also an element of $g$ (actually it is even an element of $g' =[g,g]$). 
The Ihara bracket is the Poisson bracket which arises through commutation of the derivations $D$,
\be
[D_x , D_y]  = D_{\{ x,y\} }\,,
\ee
and it therefore obeys the Jacobi identity.

By direct calculation one may compute the associator up to a given weight. We have explicitly computed it up to weight $13$ and expressed the result in terms of the same zeta-values used in the previous section. Here we will display the expansion for the associator up to weight 10,
\begin{align}
\Phi &= 1 + \zeta_2 p_2 + \zeta_3 w_3 + \zeta_4 p_4 + \zeta_5 w_5 + \zeta_2 \zeta_3 (w_2 \circ w_3) + \frac{1}{2} \zeta_3^2 (w_3 \circ w_3) + \zeta_6 p_6 \notag \\
&+ \zeta_7 w_7 + \zeta_2 \zeta_5 (p_2 \circ w_5) + \zeta_4 \zeta_3 (p_4 \circ w_3)  + \frac{1}{5} \zeta_{3,5} \{ w_5,w_3\} +\zeta_5 \zeta_3 (w_5 \circ w_3)\notag \\
&+ \frac{1}{2} \zeta_2 \zeta_3^2 ((p_2 \circ w_3) \circ w_3) + \zeta_8 p_8 + \zeta_9 w_9 + \frac{1}{6} \zeta_3^3 ((w_3 \circ w_3) \circ w_3) +\zeta_2 \zeta_7 (p_2 \circ w_7) \notag\\
&+ \zeta_4 \zeta_5 (p_4 \circ w_5) + \zeta_6 \zeta_3 (p_6 \circ w_3) + \biggl(\frac{3}{14} \zeta_5^2 + \frac{1}{14}\zeta_{3,7}\biggr) \{w_7,w_3\} + \zeta_7 \zeta_3 (w_7 \circ w_3) \notag\\
&+\frac{1}{2} \zeta_5^2 (w_5 \circ w_5) +\frac{1}{5} \zeta_2 \zeta_{3,5} (p_2 \circ \{w_5,w_3\}) + \zeta_2 \zeta_5 \zeta_3 ((p_2 \circ w_5) \circ w_3) \notag \\
&+ \frac{1}{2} \zeta_4 \zeta_3^2 ((p_4 \circ w_3) \circ w_3) + \zeta_{10} p_{10} + \ldots
\label{Phitowt10}
\end{align}
Comparing to equation (\ref{Rtowt10}) we see the obvious similarity. Going from (\ref{Phitowt10}) to (\ref{Rtowt10}), the words $p_{2r}$ are replaced by the matrices $P_{2r}$, the words $w_{2r+1}$ by the matrices $M_{2r+1}$ and the product $\circ$ by the matrix product. The antisymmetrisation of the operation $\circ$ on two elements of $g$ is the Ihara bracket which is therefore replaced by matrix commutators. 

If we consider the coproduct $\Delta_g$ on all words in the alphabet $\{e_0,e_1\}$ defined by demanding 
\begin{align}
\Delta_g (e_i) &= e_i \otimes 1 + 1 \otimes e_i\,, \\
\Delta_g (w_1 w_2) &= \Delta_g (w_1) \Delta_g (w_2)\,,
\end{align}
we find $\Phi$ is group-like,
\be
\Delta_g \Phi = \Phi \otimes \Phi\,.
\label{group-like}
\ee
Thus $\log \Phi$ is a Lie series. Moreover, since it contains no elements of length 1, it is a series in $g'=[g,g]$. 

The words $p_{2r}$ and $w_{2r+1}$ are not quite on the same footing. The reason is that all even simple zetas are related to powers of $\zeta_2$ via (\ref{evenzetas}). Hence, while all the words $w_{2r+1}$ are elements of $g'$ in accord with (\ref{group-like}), this is not the case for the words $p_{2r}$ for $r>1$. It is therefore helpful to define the words $w_{2r} \in g'$ via
\begin{align}
w_2 &= p_2\,, \notag \\
w_4 &= p_4 - \frac{1}{2 b_2} w_2 \circ w_2\,, \notag\\
w_6 &= p_6 - \frac{b_2}{b_3} w_4 \circ w_2 - \frac{1}{6 b_3} (w_2 \circ w_2) \circ w_2
\end{align}
and so on so that $\log \Phi$ is explicitly a Lie series,
\begin{align}
\log \Phi =  &\zeta_2w_2 +\zeta_3 w_3  + \zeta_4 w_4  - \frac{1}{2} \zeta_2^2 D_{w_2} w_2 + \zeta_5 w_5 + \zeta_2 \zeta_3 \Bigl(\frac{1}{2} [w_2,w_3] - D_{w_3}w_2\Bigr) \notag \\
&+ \zeta_6 w_6 + \zeta_2 \zeta_4 \Bigl( \frac{1}{2}[w_4,w_2]  - D_{w_2} w_4\Bigr) - \frac{1}{6}  \zeta_2^3 \Bigl(\frac{1}{2}[D_{w_2} w_2, w_2] - D_{w_2} D_{w_2} w_2\Bigr) \notag \\
&+ \ldots
\end{align}

To be more concrete we give the explicit form of the first few words,
\begin{align}
w_2 &= [e_1,e_0]\,\\
w_3 &= [e_0 - e_1,[e_0,e_1]]\, \\
w_4 &= -[e_0,[e_0,[e_0,e_1]]] -\frac{3}{2}[e_1,[e_0,[e_1,e_0]]] +  [e_1,[e_1,[e_1,e_0]]]\\
w_5 &= [e_0,[e_0,[e_0,[e_0,e_1]]]] - \frac{1}{2}[e_0,[e_0,[e_1,[e_0,e_1]]]] - \frac{3}{2}[e_1,[e_0,[e_0,[e_0,e_1]]]] \notag \\
&\quad + (e_0 \leftrightarrow e_1)\,.
\end{align}

We now introduce an exponentiated version of the action $\circ$, which we will denote by $\triangleleft$. Given group-like series  $A$ and $B$ in $U(g)$, we define the action of $B$ on $A$  (following \cite{Ihara,DG,fbdg}),
\be
A \triangleleft B = A(e_0 , B e_1 B^{-1}) B\,.
\ee
It is clear that the infinitesimal version of the action $\triangleleft$ defined above is $\circ$. Note that $ 1 \triangleleft A = A \triangleleft 1 = A$ and that $\triangleleft$ is associative since
\begin{align}
( A \triangleleft B ) \triangleleft C &= (A(e_0,B e_1 B^{-1}) B) \triangleleft C \notag\\
&= A(e_0,B(e_0,C e_1 C^{-1}) C e_1 C^{-1} B(e_0,C e_1 C^{-1})^{-1}) B(e_0,C e_1 C^{-1}) C\notag \\
&= A(e_0,(B \triangleleft C) e_1 (B \triangleleft C)^{-1}) (B \triangleleft C) \notag \\
&= A \triangleleft (B \triangleleft C)\,.
\end{align}

Now we are in a position to compare to the relation obeyed by the matrix $R$ in the open string amplitude. By replacing all zeta values in the expansion of $\Phi$ with their motivic versions, we obtain an element of $U(g) \otimes \mathcal{H}$ which we call $\Phi^{\rm m}$.
Calculating the coaction $\Delta$, one finds the following relation, completely analogous to (\ref{coproductrelation}),
\be
\Delta \Phi^{\rm m} = \Phi^{\rm m} \accentset{\triangleleft}{\otimes} \Phi^{\rm a}\,.
\label{Phicoproduct}
\ee
Here the symbol $\accentset{\triangleleft}{\otimes}$ indicates that the right-hand factor acts on the coefficient words of the left-hand factor via the action $\triangleleft$,
\be
\Phi^{\rm m} \accentset{\triangleleft}{\otimes} \Phi^{\rm a} = \sum_w \zeta^{\rm m}_\sha(w) \otimes w(e_0,\Phi^{\rm a} e_1 (\Phi^{\rm a})^{-1}) \Phi^{\rm a}\,.
\ee
This is a series of words with coefficients in $\mathcal{H} \otimes \mathcal{A}$.

Thus we would like to identify the matrix $R$ appearing in the open superstring amplitude with $\Phi$, with the matrix multiplication being a representation of the Ihara action $\triangleleft$.
In the next section we will show that the associator can be used to reproduce the four-point amplitude by following this logic.

\section{The four-point amplitude from the associator}
\label{sect-KZ}

The form (\ref{defassoc}) allows us to explicitly expand the associator, at least up to weights where (conjecturally) all relations between multiple zeta values are explicitly known \cite{Blumlein:2009cf}. We would like to ask if we can understand how to extract the matrices $M_{2n+1}$ and $P_{2n}$ from the form of the words $w_{2n+1}$ and $p_{2n}$. We will examine the simplest case, namely the four-point amplitude, where we will in fact be able to identify $P$ and $M$ to all orders. 

Since the matrix $R$ is a $(1 \times 1)$ matrix in the four-point case, we need a commutative realisation of the Ihara action. In other words we need a representation where the Ihara brackets $\{w_r,w_s\}$ vanish. This will guarantee that all multiple zeta values of depth greater than one will disappear. 

Now we observe that for words $w_r$ and $w_s$ in $g'$, the Ihara bracket $\{ w_r , w_s \}$ is an element of $g''=[g',g']$. This can be seen as follows. First we note that if we work modulo $g''$ then all elements of $g'$ can be written as a linear combination of the basis elements
\be
u_{kl} = {\rm ad}^k_0 {\rm ad}^l_1 [e_0,e_1]\,,
\label{ukl}
\ee
where ${\rm ad}_i x=[e_i,x]$ and the two adjoint actions commute. We recall that the Ihara bracket takes the form
\be
\{w_r , w_s \} = [w_r , w_s] + D_{w_r} w_s - D_{w_s} w_r\,.
\label{Ibr}
\ee
For $w_r , w_s \in g'$ the first term on the RHS is clearly in $g''$. On the basis elements (\ref{ukl}) we have $D_{u_{kl}} u_{k'l'} = u_{k+k'+1,l+l'+1}$ modulo $g''$ and so the final two terms on the RHS of (\ref{Ibr}) cancel modulo $g''$. Hence the Ihara brackets vanish in $g'/g''$ as required.

In this approximation, the associator in has been calculated in \cite{Drinfeld2}. For completeness we give a derivation in Appendix \ref{app-KZ}, starting from the Knizhnik-Zamolodchikov equation. The result is
\be
\log \Phi =  \frac{1}{uv}\Bigl[\frac{\Gamma(1 - u)\Gamma(1- v)}{\Gamma(1 - u - v)} -1\Bigr] [e_0,e_1], 
\ee
where $u = - {\rm ad_0}$ and $v = {\rm ad_1}$ and again we work modulo $g''$.

From the above result we deduce that
\be
\Phi = 1 + \frac{1}{uv}\Bigl[\frac{\Gamma(1 - u)\Gamma(1- v)}{\Gamma(1 - u - v)} -1\Bigr] [e_0,e_1]  \quad \text{ mod $(g')^2$.}
\ee
If we now think of $\Phi$ acting on $U(g')/(g')^2$ via the Ihara action, we find that it is represented by the multiplicative operator (i.e. a $( 1 \times 1)$ matrix),
\be
R(u,v) = \frac{\Gamma(1 - u)\Gamma(1- v)}{\Gamma(1 - u - v)}\,,
\ee
which, if we interpret the variables $u$ and $v$ as being the Mandelstam variables, is precisely the four-point amplitude $R$, with the coefficients $P_n$ and $M_n$ fixed to their correct functional forms.

We emphasise here that we have derived only the simplest superstring amplitude, the four-point one, which can of course be derived simply from many approaches. Ultimately, given the form of the worldsheet integrals describing the amplitudes, it is not surprising that it can be related to monodromies of the Knizhnik-Zamolodchikov equation \cite{varchenko}. However the approach we have outlined shows that the relation between the associator and the amplitude guarantees that the relation (\ref{coproductrelation}) holds. Moreover, one can obtain more than just the fact that the coefficients of the multiple zetas of higher depth are fixed in terms of the $M_{2n+1}$ and $P_{2n}$; one can also fix the $M_{2n+1}$ and $P_{2n}$ themselves.


\section{Closed strings and constrained multiple zeta values}
\label{sect-closed}

By emplying the Kawai-Lewellen-Tye relations \cite{Kawai:1985xq} one may obtain tree-level amplitudes for closed strings from the ordered partial amplitudes for open strings,
\be
A^{\rm closed} = (A^{\rm open})^t S A^{\rm open}\,.
\ee
Here $S$ is a $(n-3)! \times (n-3)!$ matrix consisting of particular sin factors. It obeys the important relation \cite{Schlotterer:2012ny},
\be
P^t S P = S_0,
\ee
where $S_0$ is the $\alpha' \rightarrow 0$ limit of $S$, a matrix of homogeneous rational functions of the Mandelstam variables.

The coefficients of odd zetas, the matrices $M_n$, obey (see also v3 of \cite{Schlotterer:2012ny})
\be
M_n = S_0^{-1} M_n^t S_0\,,
\label{symmetry}
\ee
implying that nested commutators $Q_{(r)} = [M_{s_1},[M_{s_2},[ \ldots [M_{s_{r-1}},M_{s_r}]\ldots]]]$ obey
\be
Q_{(r)} = (-1)^{r+1} S_0^{-1} Q_{(r)}^t S_0\,.
\ee
Since the nested commutators are the coefficients of specific multiple zeta values as dictated by (\ref{coproductrelation}), it follows that $A^{\rm closed}$ gets contributions from only certain specific combinations of multiple zeta values. For example there are no linear contributions of multiple zeta values of even depth. This has important consequences for the closed IIB superstring effective action. Since the IIB theory has an $SL(2,\mathbb{Z})$ duality symmetry, the axion and dilaton appear in the effective action through specific modular forms. At tree level these modular forms reproduce the zeta values. The restriction of the kinds of multiple zeta values appearing at tree level will have some interplay with the possible kinds of modular forms appearing. It would be interesting to investigate this point further.

In fact there are also restrictions on the kinds of multiple zeta values that can enter the open superstring amplitude. For a given number of external particles there are relations among the different commutators of the $M_n$ simply due to the fact that they are finite matrices. A very strong restriction appears at four points where, since the matrices are $(1 \times 1)$, they all commute, i.e. $[M_r,M_s]=0$. This is linked to the fact that no multiple zeta  values appear in the four-particle amplitude. At higher points the restrictions become successively weaker. For example, at five points the matrices are $(2 \times 2)$ and according to (\ref{symmetry}) above they are conjugate to symmetric matrices. This implies that the commutators $[M_r,M_s]$ are conjugate to antisymmetric matrices, but since there is a unique $(2 \times 2)$ antisymmetric matrix up to rescaling, the commutators commute with each other $[[M_r,M_s],[M_t,M_u]]=0$. Note that these constraints imply $[M_r,[M_s,[M_t,M_u]]]=[M_s,[M_r,[M_t,M_u]]]$ and that such relations were useful in \cite{Schlotterer:2012ny} in establishing that the matrix $Q$ takes an exponential form up to weight 18 in the case of the five point amplitude. Relations such as these imply that only specific linear combinations of multiple zeta values appear in the $n$-point amplitude for fixed $n$. The constraints become weaker and weaker in the sense that more and more linearly independent combinations of multiple zeta values appear as $n$ grows.

\subsection*{Acknowledgements}

JMD would like to thank the organisers of the workshop `Amplitudes and periods' held at IHES in December 2012 for the opportunity to present this work and the participants for many interesting and helpful comments. We would also like to thank Francis Brown for some helpful suggestions and insightful comments, particularly regarding the role of the associator.


\appendix

\section{Isomorphisms between $\mathcal{H}$ and $\mathcal{U}$.}
\label{}

In \cite{Schlotterer:2012ny}, the structure underlying the matrix $R$ was described by introducing an auxiliary Hopf algebra $\mathcal{U}'$, following \cite{Brown:2011ik,fb2}. This Hopf algebra is the commutative Hopf algebra of all words constructed from the alphabet $\{f_3,f_5,f_7,f_9 \ldots \}$, i.e. the alphabet with a generator of every odd degree, $r \geq 3$. The Hopf algebra product is the shuffle product $\sha$ and the coproduct is given by deconcatenation,
\be
\Delta f_{a_1}, \ldots , f_{a_n} = \sum_i f_{a_1} \ldots f_{a_i} \otimes f_{a_{i+1}} \ldots f_{a_n}\,.
\ee
By analogy with the structure of $\mathcal{H}$ above we define a trivial comodule over $\mathcal{U}'$ by
\be
\mathcal{U} = \mathbb{Q}[f_2] \otimes \mathcal{U}'\,,
\ee
where $f_2$ is of degree 2. We can think of $f_2$ as a letter which commutes with all of the $f_{2r+1}$. For convenience we will introduce the symbols $f_{2n}$ in analogy with (\ref{evenzetas}) for the even zeta values,
\be
f_{2n} = b_n f_2^n\,.
\ee

Now one may construct many isomorphisms $\phi_B$ between $\mathcal{H}$ and $\mathcal{U}$ as described in \cite{Brown:2011ik}. The map $\phi_B$ depends on the choice of basis of $\mathcal{H}$ and is constructed as follows. First one introduces an infinitesimal version of the coproduct (\ref{coproduct}) encapsulated by operators $D_{r} : \mathcal{H} \longrightarrow \mathcal{H} \otimes \mathcal{L}$ for $r$ odd and $r\geq3$ and where $\mathcal{L}$ is $\mathcal{H}$ modulo $\zeta^{\rm m}_2$ and modulo all non-trivial products. Explicitly we have
\begin{align}
D_{r} &I^{\rm m}(a_0;a_1,\ldots,a_n;a_{n+1}) = \notag \\
&\sum_{p=0}^{n-r} I^{\rm m}(a_0;a_1,\ldots,a_p,a_{p+r+1},\ldots,a_n;a_{n+1}) \otimes I^{\mathcal{L}}(a_p;a_{p+1},\dots,a_{p+r};a_{p+r+1})\,.
\end{align}
This is the coproduct $\Delta$, restricted so that the right-hand factor is taken modulo non-trivial products.

Now we construct the map $\phi_B$ by assuming that our basis contains $\zeta^{\rm m}_2$ and all the $\zeta^{\rm m}_{2n+1}$ and their products and imposing
\be
\phi_B(\zeta^{\rm m}_n) = f_n, \quad n=2,3,5,7,9,\ldots
\ee
Furthermore we impose that it is a homomorphism for the shuffle product on $\mathcal{U}'$, i.e.
\be
\phi_B(x y) = \phi_B(x) \sha \phi_B(y)\,.
\ee
Then for all non-trivial elements in the basis, i.e. the multi-zetas of higher depth, we impose recursively that
\be
\phi_B(x) =  \mu \sum_r ( \phi_B \otimes \pi_{2r+1} \circ \phi_B) D_{2r+1} (x)\,,
\ee
where $\pi_{2r+1}$ is the projection in $\mathcal{U}'_{2r+1}$ onto the word $f_{2r+1}$ of length 1 and $\mu$ is simply concatenation (not the shuffle) of words in $\mathcal{U}$, treating $f_2$ as commutative. 

Let us look at weight 8 as an example. A basis of words in $\mathcal{U}$ at weight 8 is given by $\{f_2^4, f_2 f_3 \sha f_3, f_3 \sha f_5, f_3 f_5\}$. The first three elements are the result of applying $\phi_B$ to the products $\zeta_2^4, \zeta_2 \zeta_3^2, \zeta_3 \zeta_5$. To obtain the remaining word at weight 8 we can, for example, include $\zeta_{3,5}$ in the basis $B$. Applying the operators $D_3$ and $D_5$ to $\zeta_{3,5}$ one finds
\be
D_3 \zeta_{3,5} = 0\,, \qquad D_5 \zeta_{3,5} = - 5 \zeta_3 \otimes \zeta_5\,
\ee
and hence $\phi_B(\zeta_{3,5}) = -5 f_3 f_5$. 
Having fixed the basis at a given weight, one may express all multi-zetas of that weight in terms of the basis. To decompose a particular multi-zeta value one applies the same recursive algorithm to obtain its image under $\phi_B$, except that now one allows an arbitrary amount of the unique primitive element $f_n$ of the given weight in the result. For example one finds
\be
\phi_B(\zeta_{5,3}) = 6 f_3 f_5 + f_5 f_3 + a f_8\,.
\ee
Thus we conclude that
\be
\zeta_{5,3} = -\zeta_{3,5} + \zeta_3\zeta_5 + a \zeta_8\,.
\ee
By numerical evaluation (or in this case simply application of the stuffle relation $\zeta(p) \zeta(q) = \zeta(p,q) + \zeta(q,p) + \zeta(p+q)$) one concludes that $a = -1$\,.
Had we chosen instead a different basis $B'$ where $\zeta_{5,3}$ was included as a basis element we would have found
\be
\phi_{B'} (\zeta_{5,3}) =  6 f_3 f_5 + f_5 f_3, \qquad \phi_{B'}(\zeta_{3,5}) = -5f_3 f_5 + f_8\,,
\ee
Thus $\phi_B$ and $\phi_{B'}$ define different isomorphisms between $\mathcal{H}$ and $\mathcal{U}$. Note that the first time one has an ambiguity involving an odd primitive zeta value is at weight 11 where different choices of the depth 3 element to be included in the basis result in different coefficients of $f_{11}$ in the application of $\phi_B$ to a given multi-zeta value. Many more examples on the definition and application of $\phi_B$ are given in \cite{Brown:2011ik,Schlotterer:2012ny}.

\section{Associator from the KZ equation}
\label{app-KZ}

Here we provide a derivation of the form of the associator used in section \ref{sect-KZ}. The associator can be obtained as a regularised limit as $z\rightarrow 1$ of the solution of the Knizhnik-Zamolodchikov equation,
\be
\frac{d}{d z} L(z) =  L(z) \Bigl( \frac{e_0}{z} + \frac{e_1}{1-z}\Bigr)\,,
\label{KZ}
\ee
given by a formal sum over all harmonic polylogarithms,
\be
L(z) = \sum_w w H(\tilde{w};z)\,,
\ee
where $\tilde{w}$ is the word $w$ (with $e_0$ treated as 0 and $e_1$ as 1) reversed. This reversal is needed simply because of a difference of ordering conventions between \cite{Brown:2011ik,fb2} and \cite{Remiddi:1999ew}.
The solution $L(z)$ is the unique one obeying the boundary condition,
\be
L(z) \sim z^{e_0}  \text{ as } z \rightarrow 0\,.
\ee
One can likewise define the unique solution obeying
\be
L_1(z) \sim (1-z)^{-e_1}  \text{ as } z \rightarrow 1\,.
\ee
The Drinfel'd associator can be identified with the connection relating the two solutions, 
\be
\Phi L_1(z) = L(z)\,.
\ee
From the fact that both $L$ and $L_1$ are solutions of the KZ equation and are invertible one concludes that $\Phi$ above is a constant series.

The solution $L(z)$ is divergent at $z=0$ and $z=1$ as can be seen by expanding,
\be
L(z) = 1 + e_0 H_0(z) + e_1 H_1(z) + \ldots = 1 + e_0 \log z - e_1 \log(1-z) + \ldots\,.
\ee
We regularise to obtain a quantity finite at these points,
\be
\hat{L}(z) =z^{- e_0} L(z) (1-z)^{e_1} \,.
\ee
The regularised solution obeys the following equation
\be
\frac{d}{dz} \hat{L}(z) = \hat{L}(z)  (1-z)^{-e_1}  \frac{e_0}{z} (1-z)^{e_1} - \frac{e_0}{z} \hat{L}(z) \,.
\ee
We can rewrite the first term on the RHS in terms of the adjoint action of $e_1$ on $e_0$ leading to 
\be
z \frac{d}{dz} \hat{L}(z) =  \hat{L}(z) (1-z)^{{\rm - ad}_1}(e_0) - e_0 \hat{L}(z) \,,
\label{zdzLhat}
\ee
where ${\rm ad}_i(x) = [e_i,x]$.

Now since $L(z)$ is group-like,
\be
\Delta_e L(z) = L(z) \otimes L(z)\,,
\ee
and $\hat{L}(z)$ differs from $L(z)$ only by multiplication of group-like elements $(1-z)^{e_1}$ and $z^{-e_0}$ then $\hat{L}(z)$ is also group-like. This means that it is the exponential of a Lie series in $g= {\rm Lie}[e_0,e_1]$. In fact the regularised solution $\hat{L}(z)$ actually contains no words of length 1 (they were removed by the regularisation) so we actually know that it is the exponential of a Lie series in $g'=[g,g]$,
\be
\hat{L}(z) = {\rm exp} \mathcal{L}(z), \quad \mathcal{L}(z) \in g'\,.
\ee

In the first instance we are looking for a representation of the Ihara action which is given by $(1\times 1)$ matrices of multiplicative operators. To this end we will simplify the problem and work modulo any products in $g'$ so that we can actually write
\be
\hat{L}(z) = 1 + \mathcal{L}(z) \qquad \text{ modulo products.}
\ee 
This approximation has the result that the words $w_n$ corresponding to the coefficients of the primitive elements $\zeta_n$ survive (they are elements of $g'$) but that any Ihara brackets are killed since they are all actually elements of $g''=[g',g']$. Thus we already know that no multiple zeta values will survive in this approximation.

The differential equation (\ref{zdzLhat}) above is not quite suitable for simplification since the unit term in $\hat{L}(z)$ means that the whole of the factor $(1-z)^{{\rm ad}_1}(e_0)$ contributes, even modulo products. However we can improve the situation by taking a second derivative,
\be
\Bigl(\frac{d}{dz} + z \frac{d^2}{dz^2}\Bigr) \hat{L}(z) = \frac{d}{dz} \hat{L}(z) (1-z)^{{\rm -ad}_1}(e_0)+ \hat{L}(z) \frac{{\rm ad}_1}{1-z} (1-z)^{{\rm -ad}_1}(e_0)    - e_0 \frac{d}{dz}\hat{L}(z) \,.
\ee
In the first term on the RHS we can now replace $(1-z)^{{\rm -ad}_1}(e_0)$ by $e_0$ if we work modulo products in $g'$. This term then combines neatly with the last term to $-{\rm ad}_0 \hat{L}'(z)$. In the second term on the RHS we can pull out a total adjoint action ${\rm ad}_1$ and subtract the action on $\hat{L}(z)$ and then use the first-order equation (\ref{zdzLhat}) to rewrite the result in terms of the derivative $\hat{L}'(z)$,
\begin{align}
\hat{L}(z) {\rm ad}_1 (1-z)^{{\rm -ad}_1}(e_0)  &= {\rm ad}_1 [\hat{L}(z)(1-z)^{{\rm -ad}_1}(e_0) ] - {\rm ad}_1 \hat{L}(z)(1-z)^{{\rm -ad}_1}(e_0) \,, \\
&= {\rm ad}_1 \Bigl[ z \frac{d}{dz} \hat{L}(z) + e_0 \hat{L}(z)  \Bigl] - {\rm ad}_1 \hat{L}(z) (1-z)^{{\rm -ad}_1}(e_0) \,.
\end{align}
As above, we can replace the factor $(1-z)^{{\rm -ad}_1}(e_0)$ by $e_0$ and combine the final two terms into an adjoint action of $e_0$ on ${\rm ad}_1 \hat{L}(z)$ plus a term $[e_1,e_0]$. Finally combining everything we have
\be
(1-z) z \frac{d^2}{dz^2} \hat{L}(z) +\bigl[(1-z)(1+{\rm ad_0}) - z {\rm ad}_1\bigr] \frac{d}{dz} \hat{L}(z)  - {\rm ad_0} {\rm ad}_1 \hat{L}(z) + [e_0,e_1]= 0\,.
\ee

Now, working modulo products in $g'$ means  replace the non-commuting variable $\mathcal{L}(z) = \hat{L}(z)-1$ with a function of two commuting variables ${\rm ad}_0 =-u$ and ${\rm ad}_1 = v$,
\be
\mathcal{L}(z) = \mathcal{L}(u,v,;z)[e_0,e_1]
\ee 
We can represent a general word in $U(g')$ modulo $(g')^2$ as a function of $u$ and $v$ as follows (see also the discussion around eq (\ref{ukl})),
\be
w = c_0 + \sum c_{kl} {\rm ad}_0^k {\rm ad}_1^l [e_0,e_1]   \longrightarrow c_0 + \sum c_{kl} u^{k+1} v^{l+1}\,.
\ee
In this representation the second order equation above becomes the hypergeometric equation with a constant inhomogenous term,
\be
(1-z) z \frac{d^2}{dz^2} \mathcal{L}(z;u,v) +\bigl[(1-z)(1-u) - z v \bigr] \frac{d}{dz} \mathcal{L}(z;u,v)  + u v \mathcal{L}(z;u,v)  + 1 =0\,.
\ee
The solution obeying the relevant boundary conditions is 
\be
\mathcal{L}(z;u,v) = \frac{{}_2 F_1(-u,v,1-u;z) - 1}{uv}.
\ee
The logarithm of the associator is this solution evaluated at $z=1$ (applied to $[e_0,e_1]$)
\be
\log \Phi =  \frac{1}{uv}\Bigl[\frac{\Gamma(1 - u)\Gamma(1- v)}{\Gamma(1 - u - v)} -1\Bigr] [e_0,e_1] \quad \text{ mod $g''$.}
\ee

\end{document}